# Human Behaviour as an aspect of Cyber Security Assurance

Mark Evans, Leandros A. Maglaras[*], Senior Member, IEEE, Ying He and Helge Janicke

School of Computer Science and Informatics, De Montfort University, Leicester, UK
[*]Leandros.maglaras@dmu.ac.uk

## Abstract

**There continue to be numerous breaches publicised pertaining to cyber security despite security practices being applied within industry for many years. This article is intended to be the first in a number of articles as research into cyber security assurance processes. This article is compiled based on current research related to cyber security assurance and the impact of the human element on it. The objective of this work is to identify elements of cyber security that would benefit from further research and development based on the literature review findings. The results outlined in this article present a need for the cyber security field to look in to established industry areas to benefit from effective practices such as human reliability assessment, along with improved methods of validation such as statistical quality control in order to obtain true assurance. The article proposes the development of a framework that will be based upon defined and repeatable quantification, specifically relating to the range of human aspect tasks that provide, or are intended not to negatively affect cyber security posture.**

## 1. Introduction

Information security management has grown significantly over the last 25 years and is now a common and regular item within the public domain. With buzz words such as hacking and cyber security being included within headlines and being a common topic of conversation amongst everyday technology users, information security is at the forefront of people's minds. The National Initiative for Cybersecurity Careers and Studies [1] defines cyber security within its glossary as 'The activity or process, ability or capability or state whereby information and communications systems and the information contained therein are protected from and/or defended against damage, unauthorized use or modification, or exploitation. These security-related terms have changed over the years as information security community leaders pushed the terms information security management through to information assurance (IA) up the agenda and eventually bursting into the public domain, including under its current guise of cyber security specifically addressing electronic aspects. However, the objectives have always been the same which is to primarily protect information which we process and are responsible for. Also, equally importantly there appears to be a lack of understanding within the security community as to what cyber security actually is.  For example, Health Information Trust Alliance [2] states that 'cybersecurity does not address non-malicious human threat actors, such as a well-meaning but misguided employee'.  Based on this observation this article focusses on the human factor of cyber security assurance. However, despite the huge surge in interest and acceptance of information security management, incorporating cyber security, there still appear to be gaps and weaknesses within industry and practice.  This is evident due to the large numbers of significant security incidents and data breaches that are being publicised on a regular basis including recent incidents affecting Carphone Warehouse in August 2015, TalkTalk in October 2015, Vtech in November 2015 and inadvertent email disclosure by the Bank of England in May 2015.

As a result of the continuing publication of high-profile security breaches, organisations are increasing focus [3] and looking for ways to improve their assurance in order to protect their brand and reputation, as well as to prevent or reduce the associated financial impacts [4].  This generates a picture of the inadequacy of current assurance methods for both industry and society.  Assurance techniques and approaches, in addition to technology, are required which will protect organisations and the public as a whole from continuing costly cyber security breaches.  There exist technology related breaches occurring due to malicious individuals exploiting vulnerabilities in technology on a regular basis and these are expected to continue [5] as these security hacks are now quick to appear in the media due to general public interest.  Interestingly, and perhaps surprisingly to those outside the security community, 50% of the worst breaches in the last year were caused by inadvertent human error, rising from 31% the previous year [5].  Therefore, half of significant security incidents that are occurring are due to a particular element which has not changed since the inception of information security management.  That element is people and the unintentional mistakes and errors that they make.

### 1.1. Motivation

The motivation for this article is to take a holistic look at the current status of cyber security based upon published research and recognised survey results in order to identify areas of weakness, and propose areas of further research which would advance the field of cyber security and therefore benefit wider society.  This article intends to look outside of the current practices within cyber security and identify information and research from specialised fields and industry sectors that are established and proved to be effective that could be potentially applied and assessed to understand whether positive improvements could be realised.

### 1.2. Contributions

This article makes the following contributions:
1. It reviews current materials relating to cyber security breaches, assurance research and mechanisms and publishes findings
2. Looks into the significance of the human element on cyber security assurance
3. Proposes further research using non-standard cyber security assurance mechanisms that are currently applied within other fields and highlight possible implications such as resourcing overhead.

The article from this point forward will be structured as follows. The article will look in to publicised cyber security data breaches and then move on to defining assurance and subsequently identifying current assurance methods and standards currently adopted by organisations. The document will then progress on to human factor statistics pertaining to cyber security assurance and related human behaviour that underpins these statistics. The article then moves on to mechanisms for measurement and assessment used outside of the cyber security field that could benefit the current state of cyber security based on the negative aspects earlier captured within the article.

### 2. Publicised cyber security data breaches

There have been significant volumes of serious healthcare related data breaches [6] despite the introduction of the Information Governance Toolkit (IGT) with 7255 NHS data breaches between 2011 and 2014 [7] and showing a trend of volume increases whereby there was a 101% increase from 2013 to 2014 [8]. Outside of the UK the trend continues with unintentional exposure of private or sensitive information being 83% higher for healthcare organisations than other industries but the lowest performing industry in incident response [9]. Dunn [8] also reported that 93% of breaches were due to human error and 95% of data loss in the UK is due to the cultural factors of people [10].

The UK Government 2015 security breaches survey [5] found that there had been an increase in the number of security breaches from 81% of large organisations to 90% indicating why security breaches are perceived to continue and be an expected element of business now and in the future that cannot be completely eradicated. The survey also identified that nearly 9 out of 10 large organisations surveyed now suffer some form of security breach suggesting that these incidents are now a near certainty. The report also stated that businesses should ensure they are managing the risk accordingly, and despite the increase in staff awareness training, people are as likely to cause a breach as viruses and other types of malicious software.

Interestingly the survey found that levels of security awareness delivered had gone up compared to the previous year even though staff related breaches had also risen. The survey showed that 72% of large organisations now deliver ongoing security awareness training to their staff compared with 68% the previous year. This highlights that simply pushing out standard security awareness information to the employees of an organisation is not an effective means of cyber security assurance in relation to human behaviour.

### 3. Assurance Definition

According to the National Institute of Standards and Technology [11], assurance is defined as being 'Grounds for confidence that the other four security goals (integrity, availability, confidentiality, and accountability) have been adequately met by a specific implementation. Therefore, having that in mind, it is difficult for responsible people residing at the top of the organisational hierarchy such as Chief Executive Officers, Boards, Managing Directors, Owners and Senior Managers to have confidence or guarantee that the information that their respective organisation is responsible for processing is adequately secured. This issue has been compounded by the change of terminology used over the years including utilisation of the term assurance incorrectly where it is actually referring to the underpinning controls or countermeasures being applied.

CESG [12] identified four elements of assurance within an assurance model. These four elements were intrinsic assurance, extrinsic assurance, implementation assurance and operational assurance. Based on the published cyber security incidents and breaches in the areas of operational assurance and extrinsic assurance within the field of cyber security this article will focus on those areas. CESG [12] defines operational assurance as the activities necessary to maintain the product, system or service's security functionality once it has entered operational use. Extrinsic assurance is also defined as any activity independent of the development environment which provides a level of trust in the product, system or service.

#### 3.1. Assurance Methods

There seems to be a current position within common standards whereby security assurance programmes need to be flexible [13] and require the organisation to determine what needs to be monitored and the method of monitoring as stated within clauses 9.1a and 9.1b by the British Standards Institution [14]. Standard assurance activities have been static for some time and not evolved at the pace of technology and cyber security. It is essential to have an agile security assurance framework in place to meet the needs of differing organisations and bodies. However, the current frameworks are very broad and despite being in existence for some time does not appear to be fully addressing cyber security specific assurance requirements as the breaches and statistics outlined in this article have shown.

**Human Behaviour as an aspect of Cyber Security Assurance**

According to PWC [5] the most common form of cyber risk assurance is information / cyber security risk assessment with 64% of organisations adopting this method. This position entirely relies upon the level of experience available to the organisation to interpret requirements, quantify findings effectively, develop and source assurance methods and tools, and finally communicate the cyber security status. This lack of consistency and clarity means that very few applications of cyber security assurance are the same and therefore the industry could benefit from a more prescriptive hierarchy of standards. These standards should offer greater practical guidance to organisations and providing clear quantification mechanisms for vulnerabilities associated with the human aspects of cyber security as are currently in place for technical vulnerabilities using the Common Vulnerability Scoring System (CVSS). This survey response shows that methods of assurance in relation to cyber security have not changed in order to match the current climate. Despite schemes being developed to provide assurance for Internet-facing technology such as the CESG Cyber Essentials Scheme [15] there is no wider assurance equivalent and also no published methodology addressing the assurance required relating to the human factors of cyber security. This includes clear quantification, enabling levels of cyber security effectiveness to be applied and acted upon in a consistent manner. These factors are very important as human interaction is still an essential element of cyber security despite the ever-changing technologies being made available to support assurance goals. These human activities include routine processing of electronic confidential or sensitive data through to the regular implementation and configuration of technical changes by computer system support personnel. This is a diverse range of cyber security-related activity that are essential but in isolation to not enable oversight and assurance.

Based on published scientific papers and technical reports there appears to be a heavy focus on implementing the underpinning security controls which, although essential, does not include the confirmation that that these controls have been applied correctly or as intended in order to attain assurance. This again makes the point that greater emphasis needs to be applied to assurance activities rather than just application of controls. An example of this is the McCumber cube [16] which has been, and continues to be, heavily utilised and enhanced within information security practices.

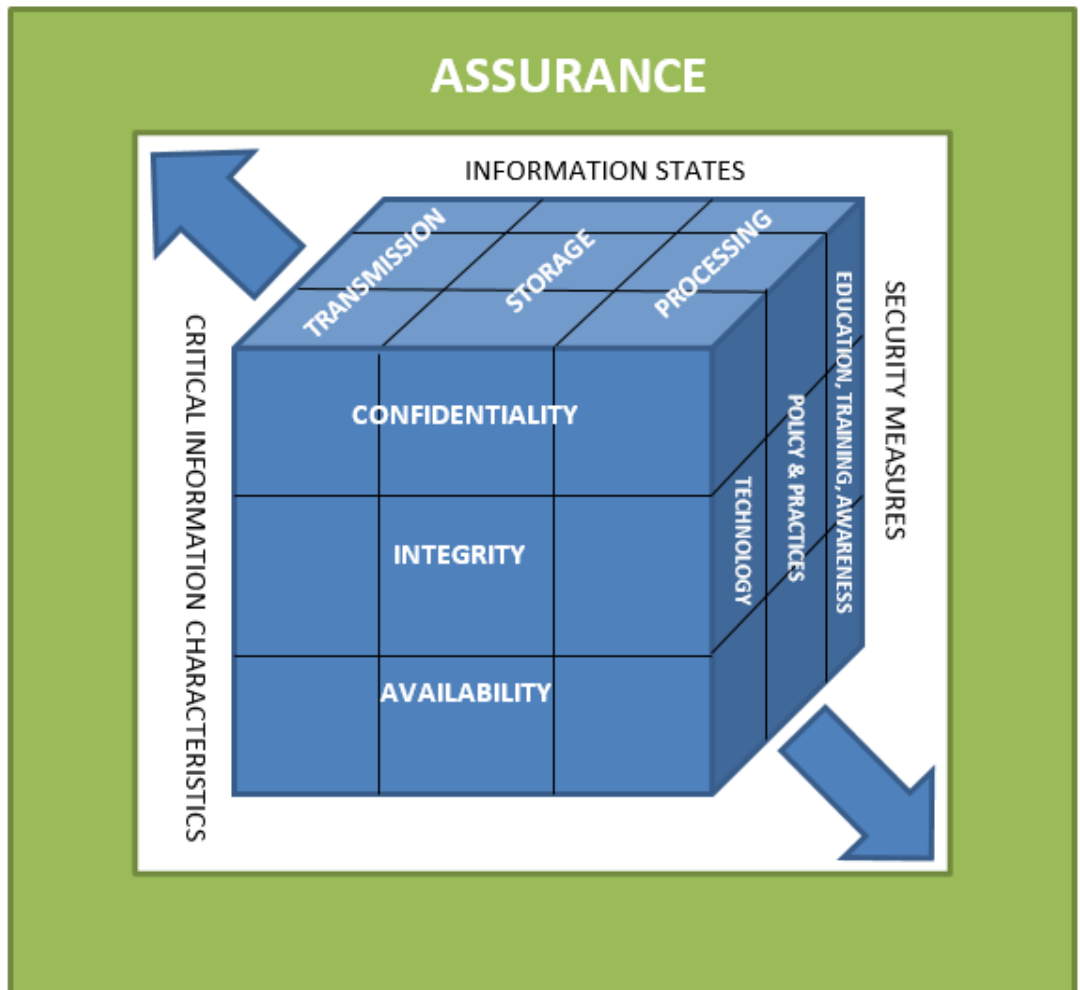

**Figure 1.** Conceptual model indicating assurance requirements encompassing the McCumber Cube

### 3.2. Common standards

There are a number of mechanisms in existence and operation currently which support cyber security assurance [17]. These include risk assessment, risk treatment, risk management, security testing and auditing. Despite these numerous mechanisms, news of high-profile security breaches are occurring and being publicised on a continuing frequent basis [18] and the impact of these breaches in financial terms doubled from 2013 to 2014 [19].

Research has shown that ISO/IEC 27001 remains the leading general standard for security management [5] and from a health perspective the key security standards underpinning the NHS Information Governance Toolkit are ISO/IEC 27001/2 [20]. Also, interestingly, the main drivers for securing sensitive data was compliance with standards [9] rather than a primary desire to protect the data for the right ethical reasons.  Research identified that 51 of the 63 Information Security Assurance requirements within the UK Health and Social Care Information Centre (HSCIC) Information Governance Toolkit originated from the ISO27000 series of standards including ISO/IEC 27001 :2013, ISO 27002:2013, 27005:2009 and associated applicable controls.  The other most prominent requirement origins included the Data Protection Act 1998, Caldicott Report and Principles, and the NHS Information Security Code of Practice.

Cherdantseva and Hilton [21] state 'an attempt to cover the entire knowledge area forces decisions to be taken that may launch a polemic'. This suggests that the current broad standards based on the principles of confidentiality, integrity and availability are too broad and therefore not effective specifically in relation to cyber security which is a view supported by the number of publicised cyber security data breaches.  Originally the ISO/IEC 27001 standard, initially known as BS 7799-2, was utilised to address business continuity planning and disaster recovery testing due to the fact that no accepted standard covering this area was available.  Now with ISO 22301, Business Continuity Management, being utilised this has allowed security professions to quite rightly focus on the security aspects of these areas rather than them both in entirety falling within the availability principle.  With other overlapping standards that can be certified against such as ISO/IEC 20000-1:2011, Information Technology – Service Management, covering security management aspects such as change management, release management, asset management and also BS 10012, Personal Information Management, used to develop a personal information management system in accordance with Data Protection legislation.  Given the current statistics captured within this article therefore  there is a need to develop a framework and hierarchy that allows whereby formal certification and assurance in relation to cyber security should be both re-scoped and also made more stringent in order to provide effective assurance.  This would also include specific assurance for the human aspect of cyber security assurance.

## 4. Current cyber security human factor statistics

There have been a number of studies and surveys undertaken relating to varying aspects of cyber security; the SANS Healthcare Cyber Security Survey [9], The Insider Threat Spotlight Report 2015 [22], Department for Business Innovation and Skills, 2014 Information Security Breaches Survey [19], and the PWC US Cybercrime survey [23] to name but a few. The Insider Threat Spotlight Report 2015 [22] stated that companies were more concerned by inadvertent insider threat data leak breaches than malicious data breaches.  However, there is no evidence of this level of concern in industry and the cyber security community in terms of change of practice.  According to the SANS Healthcare Cyber Security Survey [9], 51% considered the negligent insider as the chief threat. Yet within the 'Looking Forward' section of the document there was no mention of human security testing and in fact [23] states that only 28% of organisations are conducting employee monitoring.

The very informative PWC 2015 Information Security Breaches Survey [5] highlighted significant statistics and information pertaining to staff-related breaches which featured notably in the survey. Key findings included that three-quarters of large organisations suffered a staff-related breach and nearly one-third of small organisations had a similar occurrence, which had risen up from 58% for large organisations and 22% for small organisations compared to the previous year. These statistics show the difficulty of applying cyber security controls concerning human behaviour and interaction with confidential and sensitive information. Within larger organisations there are more processes to assure and a smaller number of information security personnel per employee. To support this finding it was also found within the survey that 72% of companies where the security policy was poorly understood had staff related breaches, which again could be down to the low ratio of information security personnel to employees to be able to clearly communicate the policy to all staff. The PricewaterhouseCoopers LLP, US cybercrime: Rising risks, reduced readiness key findings from the 2014 US State of Cybercrime Survey [23] found that for healthcare, the number of respondents who reported unintentional exposure of private or sensitive information was 83% higher than overall respondents and a critical shortcoming for a highly regulated industry that deals in sensitive personal information.

The continued evolution of technology is hugely beneficial globally and in all areas of life. However, these advances in technology, including a focus on ease of use and communication have brought with them significant changes to the cyber security landscape including broader opportunities for people within organisations at all levels access to information and also made it easier to collate, remove and circulate vast volumes of sensitive data [24] at the touch of a button with very little organisational diligence and assurance. Research found that 92% of organisations allowed access to calendar and email via mobile devices. However, 52% also allow respondents to access health records information from mobile devices [9]. It was also publicised that 15% of large organisations had a security or data breach in the last year involving smartphones or tablets which is up from 7% the previous year [5].

Whilst the internet and email has revolutionised how people communicate in the workplace, the rise of technology designed to improve collaboration, productivity and innovation has been matched by a rise of employee-related breaches affecting organisations. It was stated that communications and collaboration applications are most vulnerable to insider attack and that the perceived increase in insider attacks is due to 3 areas: awareness/training, data on mobile devices, and lack of data protection strategy or solution [22]. The PWC Information Security Breaches

Survey [5] also reported that people are the main vulnerabilities to a secure enterprise. The survey respondents believe that inadvertent human error (48%), lack of staff awareness (33%) and weaknesses in vetting individuals (17%), were all contributing factors in causing the single worst breach that organisations suffered. Regardless of the motivation of an insider, be it a deliberate act of theft or designed to embarrass an organisation; or if the breach was inadvertent due to a lack of internal controls, the threat from 'insiders' has not diminished across the UK [5].

Delving a little deeper into the statistics reveals that inadvertent human error caused half of the single worst security breaches for all respondents in 2015. This was a marked increase of over 60% year on year, and continues the trend since 2013 where accidental or inadvertent action by individuals was the main cause for the single worst breach [5]. The cyber security incidents that typically fly under the media radar are insider events. It was found that 28% of respondents pointed the finger at insiders, which includes trusted parties such as current and former employees, service providers, and contractors [23].

Although there is evidence of empirical studies that have taken place, research found that only few have been performed in terms of IT governance [25] but also further research is required relating to human behaviour and the relationship between social influence and behavioural intent [26]. Shahri, Ismail and Rahim [10] also highlighted that improving security within the healthcare organisation by adequate education and training can increase the basic knowledge and judgement of users about information security; and it can help to prevent the human errors and carelessness, but little empirical evidence supported these claims.

**5. Human Behaviour**

Research suggests that human behaviour is not consistent and can be strongly influenced by relationships, there is also a general naïve belief that bad things only happen to other people [26]. Research also found that people were willing to undertake risky practices. Individuals were actually rewarded as they were seen as helpful for allowing an event to take place without applying security controls or practice [27].

During the literature review research into other aspects of assurance and human behaviour were also investigated. These included the use of fear appeals and also user perceptions of risky behaviour pertaining to computer security. Fear appeals are persuasive communications that include an element of fear in order to receive an outcome desired by management [26]. A positive fear appeal would promote a 'danger control process' which can lead to a successful outcome as the message recipient undertakes a cognitive process to avert a threat. Fear appeals are traditionally used within healthcare and marketing such as to promote anti-smoking. Johnston and Warkentin [26] also outlined a Fear Appeals Model (FAM) incorporating components such as perceived threat severity, perceived threat susceptibility, response efficacy, self-efficacy, social influence which then leads to behavioural intent. Johnston and Warkentin [26] also states that the study aids the practice of information security management by exposing the inherent dangers of user autonomy and that end users are not consistent in their behaviours which is why a 'one-size fits all' approach to cyber security awareness and training does not offer adequate assurance. A view that is backed up by the current incident statistics highlighted earlier in this article. Also associated with the human conduct aspect of cyber security, was the undertaking of risky behaviour whereby people would undertake activity despite a known risk associated with the action. Johnston and Warkentin [26] state that individuals exhibit a rather naïve belief that bad things only happen to other people and Aytes and Connolly [27] commented that the self-image of sophisticated, security-savvy users does not track very well with their training and actual behaviours. In addition, there is a very interesting concept included by Aytes and Connolly [27] which stated: 'The vast majority of the time, users can share passwords, open e-mail attachments without checking them for viruses, and so forth, with no negative consequences. They are in fact rewarded in this behaviour, because they are either seen as helpful (in the case of sharing passwords) or they save time (by not scanning for viruses).

In relation to the fear appeal mechanism highlighted within this article, it has been shown that fear appeals [26] in isolation do not provide effective or adequate assurance, as per its definition and organisations should not rely upon this mechanism. The message could be misunderstood, forgotten or even ignored based on perceptions, relationships and social influence. Therefore, this approach should be used as an alerting mechanism only and in order to introduce assurance requires feedback to the fear appeal sender to confirm compliance. This could be a return confirmation message, scan, assessment, report, test or audit. A good analogy here would be the use of TCP in computer networking to confirm/guarantee delivery as set out later in this article. Defined assurance is essential for effective information security management as Aytes and Connolly [27] state: 'The findings suggest that it is unlikely that computer users will significantly change their behaviour in response to simply being provided with additional information regarding computing risks and practices/ and '…likely that organisations will have to enforce compliance when the risks warrant it'.

## 6. Measurement and Assessment

Metrics within cyber security is very important as it enables current state to be quantified and subsequently enable understandable and repeatable results to be communicated. It also allows organisations to understand, or set, what is or is not tolerable or acceptable. An example of this is the use of the common vulnerability scoring system (CVSS) [28] which is used to establish the severity of known technical security vulnerabilities within computer systems and software. This allows organisations, following a technical assessment, to identify the current vulnerabilities faced, confirm what is an acceptable level of exposure and address findings based on priority. However, there is no equivalent to this with regard to human behaviour within mainstream cyber security practices despite security incidents and breaches pertaining to insiders equalling those relating to external threat actors. An example of a measurement technique used within some areas of industry is statistical quality control (SQC).

Service organisations have lagged behind manufacturing firms in their use of SQC. The reason for this that SQC requires measurement and it is difficult to measure quality of a service [29] which is the primary reason why there is currently no consistent cyber security approach, quantification technique, nor associated accepted value with regard to human behaviour and its vulnerabilities. This information is essential to enable organisations to make quality decisions. For example, Rauscher and Cox [30] stated "My board has no way of knowing what we should be spending on cyber security. I could ask for 10 times as much or half of my budget". Also, "…every successful quality revolution has included the participation of upper management. We know of no exceptions". The PWC Information Security Breaches Survey [5] also found that 14% of respondents have never briefed their board on security risks, and in addition to this statistic 21% of organisations have not briefed their board in the last year showing a significant shortcoming in terms of business leaders being able to provide the assurance required as outlined earlier in this article. It was also commented that some activities, whereby direct results cannot be measured, or feedback will be delayed, rendered it ineffective as management information. An example of this could be the handling of patient identifiable information or other protected or classified material [31].

As already stated, currently within the cyber security community there are defined mechanisms for assessing threats, vulnerabilities and risks in relation to tangible aspects such as computer systems and physical environments. With regard to human behaviour the cyber security community generally appears to be accepting of the fact that there is no mainstream mechanism for assessment and quantification. However, within some industries this has been addressed through the use of human reliability assessment (HRA) and numerous underpinning techniques that have been developed. HRA involves the use of qualitative and quantitative methods to assess the human contribution to risk and has been used within high reliability industries such as petro-chemical, nuclear and aviation [32]. According to Gu et al. [33] human reliability is a term used to describe human performance such as the ability of a human to complete a given task without any errors in given conditions in a given time period. Gu et al. [33] also states that the human factors of people involved in information security can be categorised into cognition, physiology, psychology and ability and also demonstrates how incorporating HRA in to the risk assessment function significantly affects the risk assessment output. This could subsequently affect the resultant activity taken by an organisation and again emphasises the importance of reliable assurance activities and information. French et al. [34] support this view as they state that effective HRA not only complements sound technical risk analysis of the physical systems, but also helps organisations develop their safety culture and manage their overall risk. Indeed, arguably it is through this that HRA achieves its greatest effect. There are many varied methods available for HRA and one of these is called Human Error Assessment & Reduction Technique (HEART) which is a first generation HRA developed in 1985 with subsequent techniques further developed and adapted from HEART.

HEART is well validated error analysis and quantification technique [32] utilised in order to provide proactive quantification of human behaviour. It is intended to be a fast an easy method for identifying the risks associated with human error. Therefore, HEART should be a technique which is applicable to any situation or industry where human reliability is important, such as cyber security.

HEART matches the identified task to one of eight generic task categories [35]. These are:

I. Totally unfamiliar, performed at speed with no idea of likely consequences
II. Shift or restore system to a new or original state on a single attempt without supervision or procedures.
III. Complex task requiring high level of comprehension and skill.
IV. Fairly simple task performed rapidly or given scant attention.
V. Routine, highly-practiced, rapid task involving relatively low level of skill.
VI. Restore or shift a system to original or new state following procedures, with some checking.
VII. Completely familiar, well designed, highly practiced routine task occurring several times per hour, performed to highest possible standards by highly motivated, highly trained and experienced person, totally aware of implications of failure, with time to correct potential error, but without the benefit of significant job aids.
VIII. Respond correctly to system command even when there is an augmented or automated supervisory system providing accurate interpretation of system stage

The HEART process then requires the analyst or assessor to identify the applicable error producing conditions (EPC's) from a list of options ranging from 'little or no independent checking or testing of output' through to 'operator inexperience'. From this information calculations and formulae embedded within HEART are used to establish an overall human error probability (HEP) value to the identified task.  The HEART technique key elements within the quantification process are shown in Figure 2.

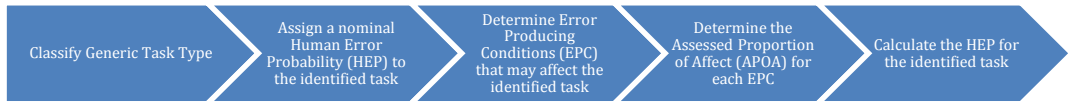

**Figure 2.** HEART quantification process

HEART is an established first generation technique for predicting human reliability and identifying ways of reducing human error.  It should be possible based on the HEART process to relatively easily apply this technique to the cyber security filed for cyber security affecting tasks performed by people.  The HEART methodology takes in to account the task and the person performing the task rather than the technical process.  However, the scope and error focus of the assessment may be too narrow [34] through the application of a first generation technique.  Second and third generation HRA techniques consider wider contexts in terms of the environment and human emotion.  The assessment must take into account the aggregated effect of people performing multiple tasks which may introduce greater likelihood of a cyber security breach or incident.

**7.  The Proposed Framework**

Due to the large number of data breach cyber security incidents, it is evident that further research needs to be undertaken to establish why such a large number of security incidents are due to human behaviour.  The lack of formal cyber security assurance relating to human behaviour set out within this article is a significant area of concern.  There is use of the term assurance but in some cases this appears to be entirely focussed upon the underpinning security controls with greater emphasis on the technical elements.  This approach doesn't provide real assurance through activities including assessment, quantification and reporting in order to provide confirmation that that these controls, including to address the risk of human error, have been applied correctly or as intended.

As shown by the publicised cyber security incidents and breaches it is evident that the current common security standards leveraged by organisations to adequately cater for human error and the associated vulnerabilities, despite current prominent reports and surveys, require enhanced focus and attention relating to human behaviour and error aspect of cyber security.  For example, one of the 35 main security categories outlined within BS ISO/IEC 27002:2013 specifically addresses technical security weaknesses (12.6 Technical Vulnerability Management) but there is no equivalent within the standard pertaining to human factor vulnerabilities. Key cyber security areas should be defined and refined through a separate modular certification approach rather than rely upon standards such as BS

ISO/IEC 27002:2013. These standards are too broad and overlap with other related standards as outlined in this article. Effective modular assurance could be achieved through separate certification for cyber security practices based on the distinct differences in current incidents and breaches.

Given the volumes of human factor related cyber security breaches and incidents, it is evident that the use of cyber security awareness training is important but organisations should consider how effective this approach is in isolation if the number of these breaches and incidents continue to increase and whether awareness alone is effective in the current climate or whether this should be enhanced through a cyber security human reliability assessment. Boards and senior management should consider whether they are taking sufficient steps to ensure a culture of strict and effective security pertaining to human error as internal, accidental factors remain the largest cause of cyber security breaches [5].

A technology scenario where return confirmation rather than a one-way communication, as a form of assurance, has been applied is the use of the Transmission Control Protocol (TCP) within TCP/IP (Internet Protocol) computer networking. TCP is one of the main protocols in TCP/IP networks. Whereas the IP protocol deals only with packets, TCP enables two hosts to establish a connection and exchange streams of data. TCP guarantees delivery of data and also guarantees that packets will be delivered in the same order in which they were sent. Without the use of TCP there would just be an assumption that the connection had been successfully established and data delivered. Another computer networking protocol which does not undertake message receipt confirmation is the User Datagram Protocol (UDP); a [connectionless](#) [protocol](#) which provides a direct way to send and receive data and is used primarily for [broadcasting](#) messages over a computer network. Therefore, a cyber security analogy using these protocols would be:

- TCP - Security Manager of an organisation sends out an email alert to staff asking them to remove all client personal data from their desktop computer hard drives and store it on networked file servers where the data is secured and backed up on a regular basis. However, the Security Manager also asks the message recipients to acknowledge receipt and understanding of the instruction and confirm when the task has been completed.
- UDP – Security Manager of an organisation sends out a broadcast email alert to staff asking them to remove all client personal data from their desktop computer hard drives and store it on networked file servers where the data is secured and backed up on a regular basis.

Obviously the UDP form of confirmation is much quicker and requires less management and interaction, however the Security Manager in the scenario above would not be able to provide assurance that the task had been completed or even the instruction received by the intended recipients. Whereas, the TCP form would require confirmation to be sent to the Security Manager that would allow a greater degree of assurance that the instruction had been received by the intended recipient, understood, and that the staff members believe they have complied with the requirement. In order to attain full assurance a form of independent testing would need to be undertaken with results checked and communicated. Only then could the organisation really provide assurance that all personal data had been moved on to the central server as per the instruction.

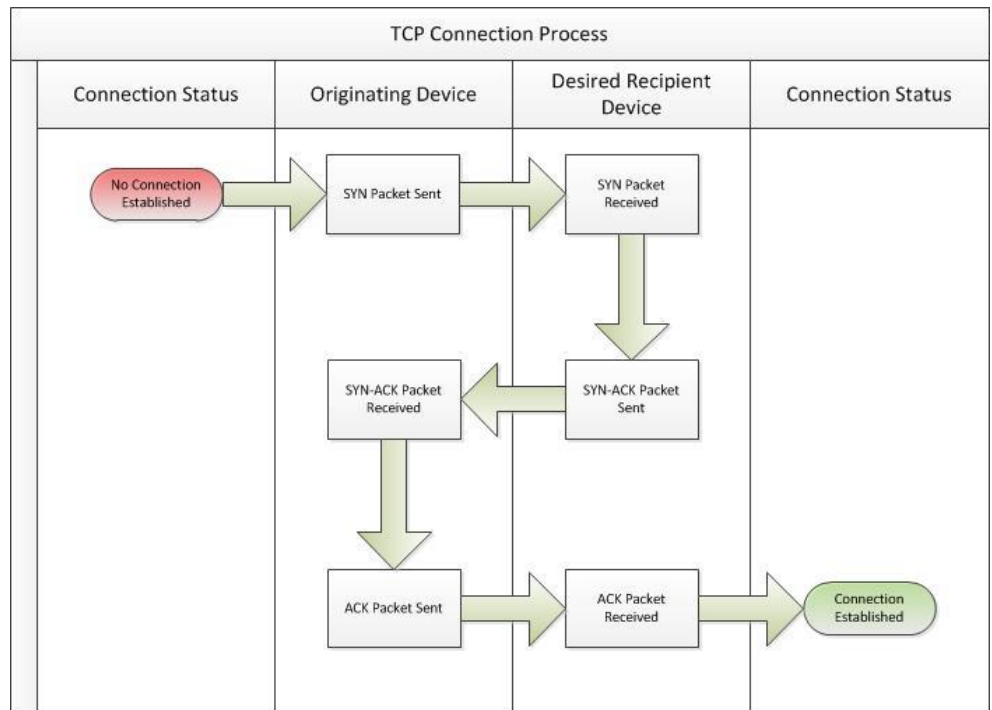

**Figure 3.1**. TCP Connection Process

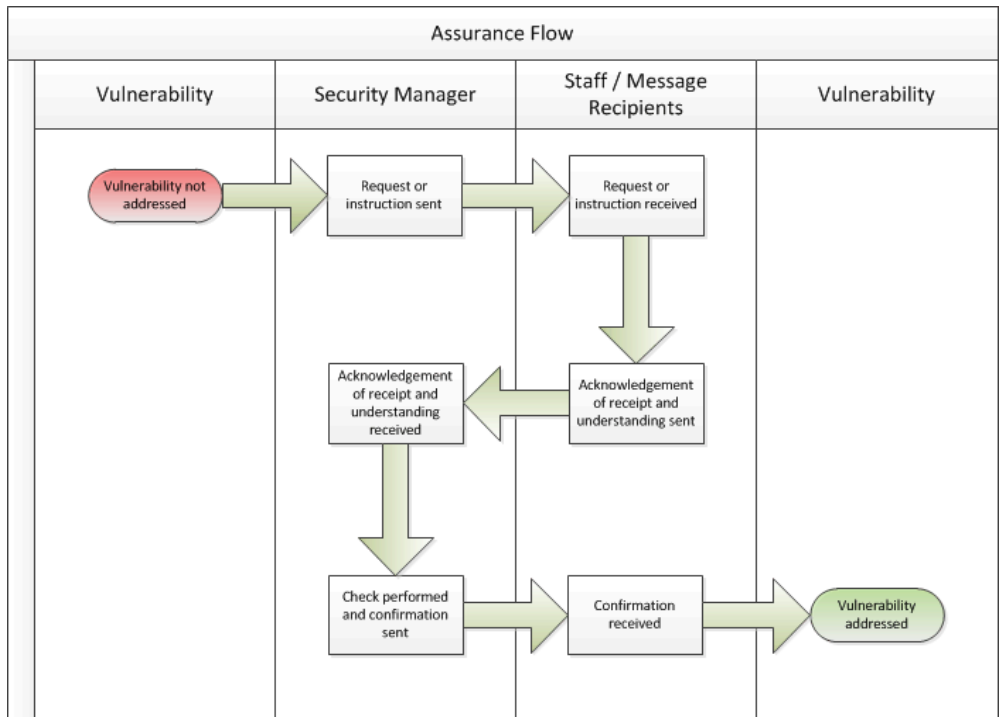

**Figure 3.2**. Assurance process

Organisations generally now understand that they should be measuring and monitoring their security controls through common channels now such as penetration testing, vulnerability assessment, risk assessment, audit, patching reports, incident statistics, and anti-virus software updates and coverage with internal audit and information/cyber risk assessment being the most common [5]. These forms of assurance are definitely required and essential but when the management information they are providing is analysed they are ultimately retrospective and technology focussed. Therefore, the human error with regard to cyber security has been found to be either too difficult, not able to provide financial reward, or felt to be not required despite the headlines we are often faced with. Human reliability assessment methodologies and techniques have been developed and implemented within other industries but not cyber security to date. Methodologies such as HEART were developed approximately 20 years ago but still have not flowed in to mainstream information security practice in addition to the use of formal measurement techniques such as Statistical Quality Control (SQC) which are common in manufacturing environments [29]. The difficulties of quantifying human reliability within cyber security have not been developed and is not currently within mainstream information security practice. The cyber security community should include a greater focus on quantification of all areas, including human reliability, to provide clear quality management information to Boards and senior management within organisations which in turn will allow greater cyber security assurance to be attained.

With some technique modification, including adjusting evaluation focus to the human error potential pertaining to the use of, and concurrent access to, multiple data stores and applications which could result in cyber security incidents and breaches, the adaption of HEART or another HRA technique could benefit the cyber security community.

In order to provide real assurance there must be a cyclic or return flow of information between the instigator and elements within scope of the activity, but in many cases this is not the case. For example, in order to provide assurance an instruction must be communicated, a change implemented, a form of check undertaken, and the results of the check confirmed against the initial instruction as can be seen below in Figure 4.

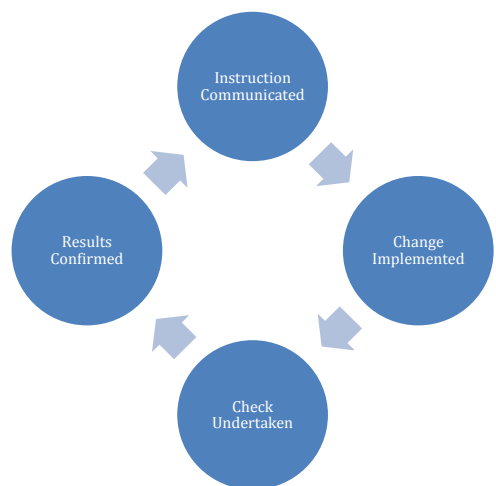

**Figure 4**. Basic Assurance Cycle

This article has shown that a defined assurance model is required that interconnects with the models already developed such as the McCumber cube in order to enhance cyber security as shown in Figure 1. Following the literature review an assurance framework is proposed that integrates human reliability assessment, statistical quality control and a vulnerability scoring system that pertains to human rather than technical vulnerabilities. The framework should also be suitable for all cyber security affecting tasks performed by humans from routine processing of personal data through to technical application of security updates by administrative personnel. Further research and development in this area should be undertaken, and as well as looking at the actual framework should also research methods of completion that could potentially ease the resourcing burden associated with the task. Based on an organisational compliance program associated with human error related to the protection of electronic systems and data tests are proposed to establish the comparable accuracy of the assessment being performed by a security professional, non-security personnel and also employee self-assessment. A conceptual high-level model of the proposed assurance framework is shown in Figure 5.

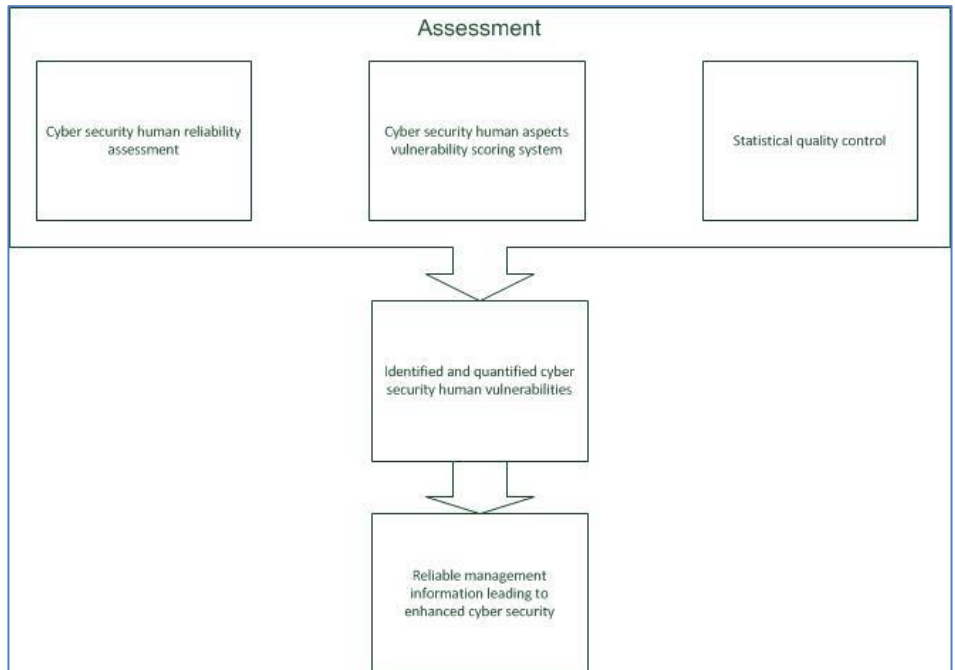

**Figure 5**. Proposed high-level cyber security human vulnerability model

There is an overhead associated with the proposed assurance framework in that organisations are required to invest in applying greater resources and time to meeting assurance requirements. This will therefore require greater focus, attention and expenditure within cyber security assurance as this framework is not looking to introduce efficiencies but to enhance effectiveness which at this time comes with increased resource obligations. These resource requirements could come from internal resources or be external independent resources as undertaken currently as part of technical security testing techniques.

**8. Conclusion**

As outlined within this article, organisations and society continue to be affected by both regular and similar cyber security breaches. These breaches pertain to technical implementations as well as routine processing of confidential electronic information. Despite this range of activities, it has been proven that half of these have human error at their core. Therefore, there should be increased empirical and theoretical research in to human aspects of cyber security based on the volumes of human error related incidents in order to establish ways in which mainstream cyber security practice can benefit.

This article has demonstrated that there is further research required in to cyber security assurance and quantification in relation to human factors to develop an effective assurance framework. This approach would benefit the field of cyber security as a common useable solution is not currently available and organisations are relying upon independent skills and knowledge of individuals. It is proposed that a specific framework is developed based upon defined and repeatable quantification specifically relating to the range of human aspect tasks that provide, or are intended not to negatively affect cyber security posture. Techniques that this framework should be built upon include human reliability assessment, statistical quality control and a cyber security human aspect vulnerability scoring system. In conclusion, the cyber security community should continue to progress and develop but it must not forget its roots and the obvious statistics that indicate we have not yet addressed the risks associated with the one consistent element of cyber security, the human error.

Human Behaviour as an aspect of Cyber Security Assurance